\documentclass[aps,prc,twocolumn,showpacs,superscriptaddress,longbibliography,floatfix,10pt]{revtex4-2}
\usepackage{indentfirst}
\usepackage{bm}
\usepackage[dvipsnames]{xcolor}
\usepackage{dcolumn}
\usepackage{amssymb}
\usepackage{amsmath}
\usepackage{graphicx}
\usepackage{dcolumn}
\usepackage{bm}
\usepackage{xcolor}
\usepackage{fix-cm}
\usepackage{mathptmx}
\usepackage[T1]{fontenc}
\usepackage{epstopdf}
\usepackage{longtable}
\usepackage{ulem}
\usepackage[colorlinks,citecolor=blue,linkcolor=blue,anchorcolor=blue,filecolor=blue,urlcolor=blue]{hyperref}
\usepackage{multirow}
\makeatletter
\def\NAT@def@citea{\def\@citea{\NAT@separator}}
\makeatother

\begin{document}	
	\hyphenpenalty=5000
	\tolerance=2000
	
        \title{Potential signature of new magicity from universal aspects of nuclear
    charge radii}
  
		\author{Dan Yang}
		\affiliation{Department of Physics, Guangxi Normal University, Guilin, 541004, China}
  		\affiliation{Guangxi Key Laboratory of Nuclear Physics and Technology, Guilin, 541004, China}

		\author{Yu-Ting Rong}
		\email[Corresponding author:~]{rongyuting@gxnu.edu.cn}
		\affiliation{Department of Physics, Guangxi Normal University, Guilin, 541004, China}
  		\affiliation{Guangxi Key Laboratory of Nuclear Physics and Technology, Guilin, 541004, China}

		\author{Rong An}
		\email[Corresponding author:~]{rongan@nxu.edu.cn}
		\affiliation{School of Physics, Ningxia University, Yinchuan 750021, China}
  		\affiliation{Guangxi Key Laboratory of Nuclear Physics and Technology, Guilin, 541004, China}
        \affiliation{Key Laboratory of Beam Technology of Ministry of Education, School of Physics and Astronomy, Beijing Normal University, Beijing 100875, China}

            \author{Rui-Xiang Shi}
		\affiliation{Department of Physics, Guangxi Normal University, Guilin, 541004, China}
  		\affiliation{Guangxi Key Laboratory of Nuclear Physics and Technology, Guilin, 541004, China}

		\date{\today}
		
		\begin{abstract}

       Shell quenching phenomena in nuclear charge radii are typically observed at the well-established neutron magic numbers. However, the recent discovery of potential new magic numbers at the neutron numbers $N = 32$ and $N = 34$ has sparked renewed interest in this mass region. This work further inspects into the charge radii of nuclei around the $N = 28$ shell closure using the relativistic Hartree-Bogoliubov model. We incorporate meson-exchange and point-coupling effective nucleon-nucleon interactions alongside the Bogoliubov transformation for pairing corrections. To accurately capture the odd-even staggering and shell closure effects observed in charge radii, neutron-proton correlations around Fermi surface are explicitly considered. {The charge radii of Ca and Ni isotopes are used to test the theoretical model and show an improvement with neutron-proton pairing corrections, in particular for neutron-rich isotopes.}
        Our calculations reveal a inverted parabolic-like trend in the charge radii along the $N = 28$ isotones for proton numbers $Z$ between 20 and 28. Additionally, the shell closure effect of $Z = 28$ persists across the $N = 28$, $30$, $32$, and $34$ isotonic chains, albeit with a gradual weakening trend. Notably, the significantly abrupt changes in charge radii are observed across $Z = 22$ along both the $N = 32$ and $N = 34$ isotonic chains. {This kink at $Z = 22$ comes from the sudden decrease of the neuron-proton correlation around Fermi surfaces across $Z = 22$ for $N = 30$, 32, and 34 isotones, and might provide a signature for identifying the emergence of neutron magic numbers $N=32$ and 34.}
        Furthermore, the calculated charge radii for these isotonic chains ($N = 28$, 30, 32, and 34) can serve as reliable guidelines for future experimental measurements.

		\end{abstract}
		
		
		\maketitle
	
\section{INTRODUCTION}

Charge radius is one of the directly measured properties related to nuclear structure phenomena, such as
halo structure~\cite{Mueller2007_PRL99-252501,Geithner2008_PRL101-252502,Noertershaeuser2009_PRL102-062503}, odd-even staggering (OES)~\cite{Barzakh2017_PRC95-014324,Hammen2018_PRL121-102501,Sels2019_PRC99-044306,Barzakh2021_PRL127-192501,DayGoodacre2021_PRL126-032502}, 
 and shape-phase transition~\cite{Marsh2018_NP14-1163, Peru2021_PRC104-024328}. Meanwhile, the available charge radii data are useful for pinning down the isovector components in the equation of state of isospin asymmetric nuclear matter~\cite{Wang2013_PRC88-011301,Pineda2021_PRL127-182503,Brown2020_PRR2-022035,Novario2023_PRL130-032501,An2023_NST34-119}. Owing to the advanced techniques in experiment~\cite{Campbell2016_PPNP86-127,Yang2023_PPNP129-104005}, {more charge radii of finite nuclei including those far away from the $\beta$-stability line are measured~\cite{Angeli2013_ADNDT99-69,Li2021_ADNDT140-101440}.} These highly precise databases can be employed to encode the unknown information on nucleon-nucleon interactions and complex dynamics of protons and neutrons moving inside the nucleus.

Nuclear size can be ruled by the $A^{1/3}$~\cite{Bohr1969} or $Z^{1/3}$~\cite{Zhang2002_EPJA-285} law.
The validity of these empirical formulas is challenged in describing the discontinuity changes of nuclear charge radii. 
In particular, the shell quenching phenomena of nuclear charge radii can be observed naturally at the traditional neutron magic numbers~\cite{Vermeeren1992_PRL68-1679,Cocolios2011_PRL106-052503,Kreim2014_PLB731-97,Gorges2019_PRL112-192502}, namely the shrunken trend of charge radii at the fully filled shells. Conversely, the shrunken trend of the charge radii along a long isotopic chain provides an alternative signature to identify the shell closure effect. Recent studies suggest that the new magicity of $N=32$ and $N=34$ can be found in some specific isotopic chains, such as in the Ar~\cite{Kristian2023_PRL131-102501}, K~\cite{Rosenbusch2015_PRL114-202501}, Ca~\cite{Wienholtz2013_Nature498-346,Gallant2012_PRL109-032506,Michimasa2018_PRL121-022506}, Sc~\cite{Xu2019_PRC99-064303}, and Ti~\cite{Janssens2002_PLB546-55,Leistenschneider2018_PRL120-062503} isotopes. 
However, the unexpectedly large charge radius for $^{52}$Ca has been detected beyond the fully filled $N=28$ shell closure~\cite{GarciaRuiz2016_NP12-594}. The same scenario can also be encountered in K isotopes, in which the abrupt change of charge radius is not emerged from $^{51}$K to $^{52}$K~\cite{Koszorus2021_NP17-439}. As suggested in Refs.~\cite{Kortelainen2022_PRC105-L021303,An2024_PRC109-064302}, nuclear charge radii between neutron numbers $N = 28$ and $N = 40$ exhibit a universal pattern, which is independent of the proton number in a nucleus. This seems to reveal that it is difficult to identify the new magicity of the neutron numbers $N=32$ and $N=34$ from the characteristics of nuclear charge radii along a long isotopic chain.

As is well known, charge radii
are influenced by various underlying mechanisms, such as
pairing correlations~\cite{Reinhard2017_PRC95-064328,Perera2021_PRC104-064313,Miller2019_PLB793-360,Souza2020_PRC101-065202,Cosyn2021_PLB820-136526}, shape evolutions~\cite{Lalazissis1996_NPA597-35,An2023_CTP75-035301}, 
cluster structures~\cite{Mueller2007_PRL99-252501,Geithner2008_PRL101-252502}, shell evolutions~\cite{Barzakh2018_PRC97-014322,Nakada2019_PRC100-044310,DayGoodacre2021_PRL126-032502}, and center-of-mass correlations~\cite{Long2004_PRC69-034319,Reinhard2021_PRC103-054310,Rong2023_PRC108-054314}.
To further elucidate the behaviour of charge radii for nuclei with neutron numbers beyond $N=28$, a reliable and unified method should be employed in describing the systematic evolution of nuclear charge radii. 
{The Fayans energy density functional (EDF) model which takes into account the components of pairing interactions can well reproduce the local variations of nuclear charge radii along even-$Z$ isotopic chains~\cite{Reinhard2017_PRC95-064328}.} For odd-$Z$ cases, the obviously overestimated amplitudes of OES in charge radii are observed ~\cite{Groote2020_NP16-620,Koszorus2021_NP17-439}. The same scenarios also occur in the modified relativistic mean-field plus Bardeen-Cooper-Schrieff (BCS) equation ansatz, namely the RMF(BCS)* model~\cite{An2020_PRC102-024307,An2022_CPC46-054101,An2022_CPC46-064101}.
In this approach, the neutron-proton pairing correlations around Fermi surface have been incorporated into the root-mean-square (rms) charge radii formula. The RMF(BCS)* model can also reproduce the OES and shell closure effects of charge radii well. Recently, a revised version has been proposed by introducing the correlation of the simultaneously unpaired neutron and proton around Fermi surface~\cite{An2024_PRC109-064302}. In this revised version, along a long isotopic chain, the abrupt increases in charge radii donot found around $N=32$ and $34$ isotopes. A rapid trend of changes of charge radii across $Z=22$ is observed along $N=32$ and $N=34$ isotones, which may provide a sensitive indicator to reveal the emergency of new magicity from the systematic evolution of nuclear charge radii.

However, in their calculations, the pairing correction is considered by the BCS approximation, which is not suitable for nuclei far away from the $\beta$-stability line~\cite{Dobaczewski1984_NPA422-103,Dobaczewski1996_PRC53-2809}. For example, in Ref.~\cite{Miller2019_NP15-432}, the authors found that Fy($\Delta r$, HFB) is better than Fy($\Delta r$, BCS)~\cite{Reinhard2017_PRC95-064328} in describing the charge nuclei of $^{36-38,52}$Ca. Another case is that only the meson-exchange forces are used in their work. The results with point-coupling forces, which are widely used since they can resolve short-distance dynamics in low energies and easy to extend to excited states~\cite{Nikolaus1992_PRC46-1757,Buervenich2002_PRC65-044308,Niksic2006_PRC73-034308,Niksic2006_PRC74-064309,Niksic2008_PRC78-034318,Zhao2010_PRC82-054319,Yao2011_PRC83-014308,Rong2023_PLB840-137896}, should be reviewed.
Therefore, in order to further investigate the influence of neutron-proton correlations around Fermi surface on determining the charge radii, the multidimensionally-constrained relativistic Hartree- Bogoliubov (MDC-RHB) model~\cite{Zhao2015_PRC92-064315,Zhou2016_PS91-063008} is employed in this work. 
In this model, both the meson-exchange and point-coupling nucleon-nucleon effective interactions can be taken into account, and the pairing correlation is tackled by Bogoliubov transformation under a separable pairing force of finite range.

This work is organized as follows. In Sec.~\ref{sec2}, we introduce the MDC-RHB model and the neutron-proton pairing correlation extract from the quasi-particle states around Fermi surface. In sec.~\ref{sec3}, {the charge radii for Ca and Ni isotopes are examined, and those for $N=28$, $30$, $32$, and $34$ isotones} are investigated with both meson-exchange and point-coupling effective interactions, followed by a summary in Sec.~\ref{sec4}.

\section{THEORETICAL FRAMEWORK}\label{sec2}


In the MDC-RHB model, the RHB equation in coordinate space can be recalled as follows~\cite{Ring1996_PPNP37-193,Kucharek1991_ZPA339-23}:
\begin{equation}
	\label{eq:rhb}
	\int d^{3}\bm{r}^{\prime}
	\left( \begin{array}{cc} \bm{h}-\lambda  &  \Delta                      \\
		-\Delta^{*}   & -\bm{h}+\lambda \end{array}
	\right)
	\left( \begin{array}{c} U_{k} \\ V_{k} \end{array} \right)
	= E_{k}
	\left( \begin{array}{c} U_{k} \\ V_{k} \end{array} \right),
\end{equation}
where $\lambda$ is the Fermi energy, $\Delta$ is the pairing tensor field, $E_k$ is the quasi-particle energy, and $(U_k(\bm{r}),V_k(\bm{r}))^T$ is the quasi-particle wave function.
$\bm{h}$ is the single-particle Hamiltonian which can be written as follows:
\begin{equation}
	\bm{h} = \bm{\alpha} \cdot \bm{p}  +
	\beta \left[ M + S(\bm{r}) \right]+ V(\bm{r})+\Sigma_R(r),
\end{equation}
where $M$ is the mass of nucleon, $S(\bm{r})$, $V(\bm{r})$, and $\Sigma_R(r)$ are the scalar, vector, and rearrangement potentials, respectively.
For meson-exchange interactions,
\begin{equation}
	\begin{aligned}
		&S(\bm{r})=g_\sigma \sigma, \\
		&V(\bm{r})=g_\omega \omega_0+g_\rho \rho_0 \cdot \tau_3+e\dfrac{1-\tau_3}{2}A_0, \\
		&\Sigma_R(\bm{r})=\dfrac{\partial g_\sigma}{\partial \rho_V} \rho_S \sigma + \dfrac{\partial g_\omega}{\partial \rho_V} \rho_V \omega_0 + \dfrac{\partial g_\rho}{\partial \rho_V} \rho_V\tau_3 \rho_0,
	\end{aligned}  
\end{equation} 
where $g_\sigma$, $g_\omega$, and $g_\rho$ are coupling constants of $\sigma$, $\omega_0$, and $\rho_0$ meson fields, $A_0$ is the time-like component of the Coulomb field mediated by photons naturally, $e$ is the charge unit for protons. $\tau_3=1$ and $-1$ for neutron and proton, respectively. $\rho_S$ and $\rho_V$ are isoscalar and isovector densities, respectively.
For point-coupling interactions,
\begin{equation}
	\begin{aligned}
		S(\bm{r})=&\alpha_S\rho_S+\alpha_{TS}\rho_{TS}\tau_3 + \beta_S	\rho_S^2 +\gamma_S\rho_S^3 \\
		&+\delta_S\Delta \rho_S +\delta_{TS}\Delta \rho_{TS}\tau_3, \\
		V(\bm{r})=&\alpha_V\rho_V+\alpha_{TV}\rho_V\tau_3 +\gamma_V\rho_V^3 \\
		&+\delta_V\Delta\rho_V+\delta_{TV}\Delta \rho_{TV}\tau_3 +e\dfrac{1-\tau_3}{2}A_0, \\
		\Sigma_R(\bm{r})=&\dfrac{1}{2} \dfrac{\partial \alpha_S}{\partial \rho_V} \rho_S^2 
		+ \dfrac{1}{2} \dfrac{\partial \alpha_V}{\partial \rho_V} \rho_V^2 +\dfrac{1}{2} \dfrac{\partial \alpha_{TV}}{\partial \rho_{V}} \rho_{TV}^2,
	\end{aligned}  
\end{equation}
where $\alpha_S$, $\alpha_V$, $\alpha_{TS}$, $\alpha_{TV}$, $\beta_S$, $\gamma_S$, $\gamma_V$, $\delta_S$, $\delta_V$, $\delta_{TS}$, and $\delta_{TV}$ are coupling constants for different channels, $\rho_{TS}$ and $\rho_{TV}$ are time-like components of isoscalar current and time-like components of isovector current, respectively.
The pairing field $\Delta$ is calculated by the effective pairing interaction $V$ and the pairing tensor $\kappa$ as follows:
    \begin{equation}
	      \begin{aligned}
	    	&\Delta(\bm{r}_{1}\sigma_{1},\bm{r}_{2}\sigma_{2}) 
	    	= 
	    	\int d^{3}\bm{r}_{1}^{\prime} d^{3}\bm{r}_{2}^{\prime} 
	    	\sum_{\sigma_{1}^{\prime}\sigma_{2}^{\prime}}  \\
	    	&	V(\bm{r}_{1}         \sigma_{1},          \bm{r}_{2}         \sigma_{2},
	    	\bm{r}_{1}^{\prime}\sigma_{1}^{\prime}, \bm{r}_{2}^{\prime}\sigma_{2}^{\prime}) 
	    	\kappa 
	    	(\bm{r}_{1}^{\prime}\sigma_{1}^{\prime}, 
	    	\bm{r}_{2}^{\prime}\sigma_{2}^{\prime}).
	      \end{aligned}
       \end{equation}
 In this work, a separable pairing force of finite
range with pairing strength $G = 728$ MeV fm$^3$ and effective
range of the pairing force $a = 0.644$ fm are adopted~\cite{Tian2009_PLB676-44}.

The modified rms charge radii formula is recalled as follows (in unit of fm$^2$)~\cite{An2024_PRC109-064302}:
\begin{equation}\label{eq:r_ch}
 r_{\rm ch}^2=\langle r_p^2 \rangle +0.7056~{\rm fm}^2+\dfrac{a_0}{\sqrt{A}}\Delta D~{\rm fm}^2 +\dfrac{\delta}{\sqrt{A}}~{\rm fm}^2.
\end{equation}
where the first term represents the charge distribution of point-like protons, and the second term is incorporated due to the finite
size of protons, $A$ is the mass number and ${a_0}$ is a normalization constant. The term $\Delta D=|D_n-D_p|$ is employed to measure the neutron-proton correlations around Fermi surface, which reads: 
\begin{equation}\label{eq:Dnp}
    D_{n,p}=\sum_{k>0} u_k^{n,p} \upsilon_k^{n,p},
\end{equation}
where $u_k^{n,p}$ is the occupation probability of the $k$th quasi-particle orbital for neutron or proton, and $\upsilon_k^2={1-u_k^2}$. In practice, the quasi-particle levels that satisfy $|E_k-\lambda|<20$ MeV are summed up in Eq.~(\ref{eq:Dnp}). The
values of $a_0=$0.561 and $\delta=$0.355(0.000) for odd-odd (even-even, odd-even, and even-odd) nuclei are the same as those in Ref.~\cite{An2024_PRC109-064302}. 

In the MDC-RHB model, one can allow for mutipole moments under $V_4$ symmetry. In this work, we restrict the calculations into both axial and reflection symmetries, namely, only quadrupole deformation $\beta_{20}$ is considered. The effective interactions NL3~\cite{Lalazissis1997_PRC55-540} and PC-PK1~\cite{Zhao2010_PRC82-054319} are used in our calculations.

\section{RESULTS AND DISCUSSIONS}\label{sec3}

\begin{figure*}[htbp!]
\centering
\includegraphics[width=0.35\textwidth]{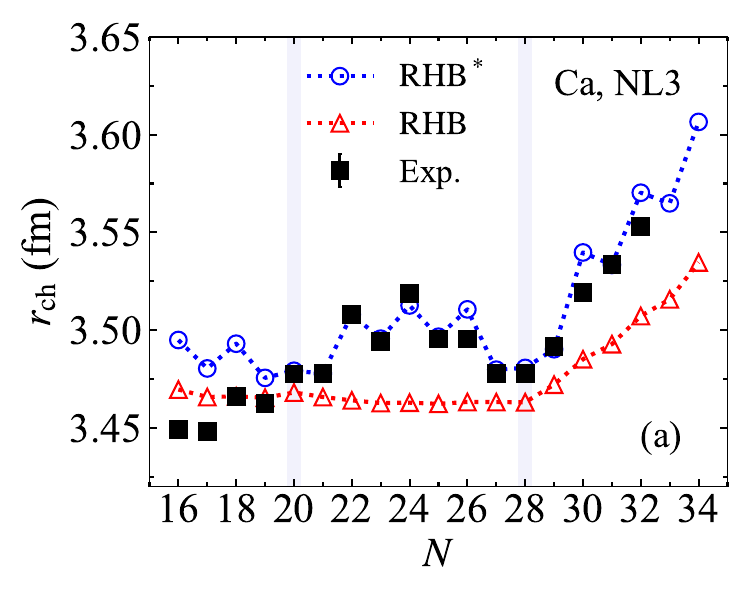}
\includegraphics[width=0.35\textwidth]{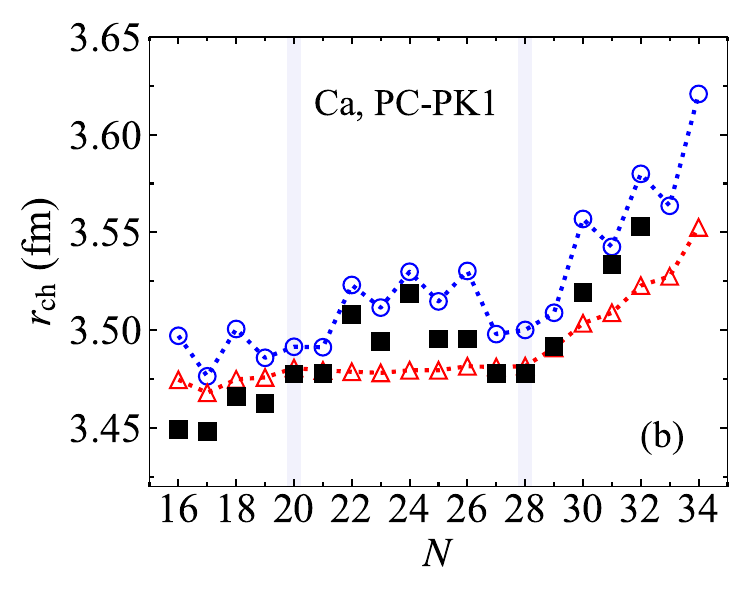}
\includegraphics[width=0.35\textwidth]{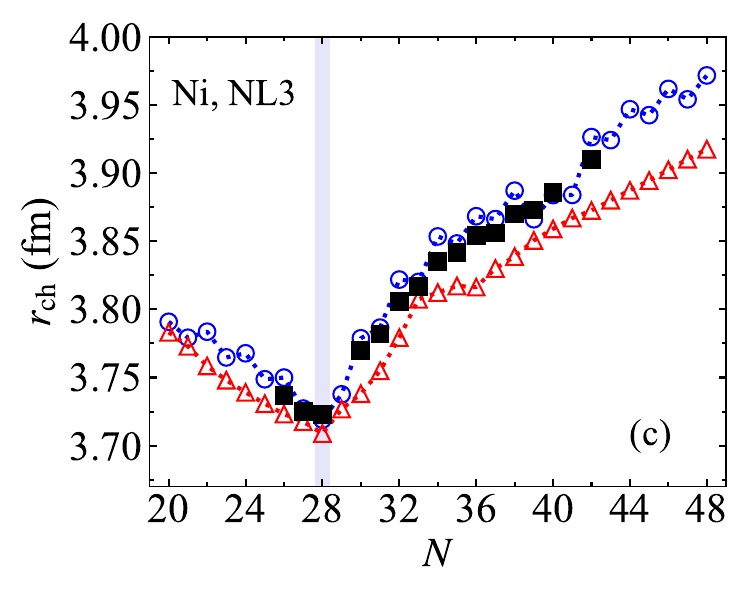}
\includegraphics[width=0.35\textwidth]{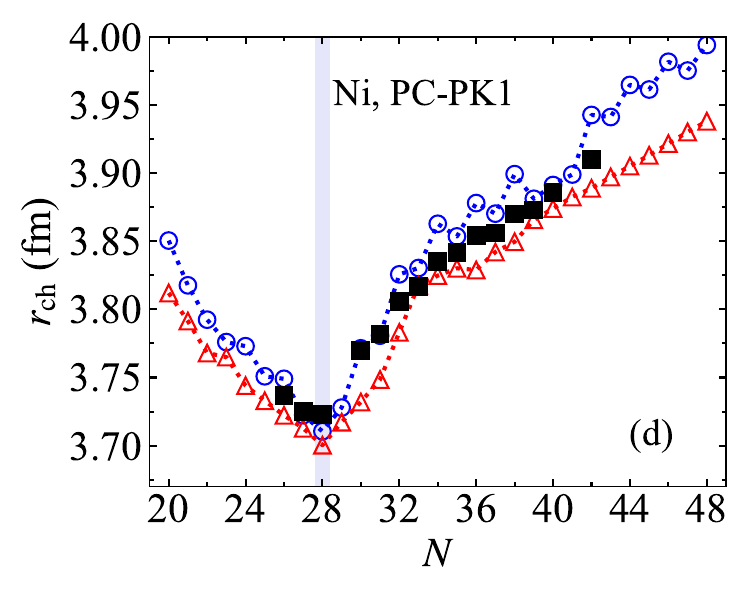}
\caption{(Color online) Charge radii of Ca isotopes calculated with (a) NL3 and (b) PC-PK1, and Ni isotopes calculated with (c) NL3 and (d) PC-PK1. The results calculated by RHB* (RHB) model are shown by open circles (open triangle). The experimental data (solid
square) are taken from Refs.~\cite{Angeli2013_ADNDT99-69,Li2021_ADNDT140-101440,Sommer2022_PRL129-132501,Malbrunot-Ettenauer2022_PRL128-022502}. The gray band represents the proton numbers $N = 20$ and $N=28$.}\label{fig0}
\end{figure*}

{The charge radii of Ca and Ni isotopes are typical examples for discussing $N=20$ and $N=28$ shell closures and testing theoretical models~\cite{GarciaRuiz2016_NP12-594,Sommer2022_PRL129-132501,Malbrunot-Ettenauer2022_PRL128-022502}. Therefore, we first calculate the charge radii of them with the MDC-RHB model. For the sake of discussion, the results obtained with and without the neutron-proton correlations around Fermi surface are marked by the RHB* and the RHB, respectively.
Fig.~\ref{fig0}(a) and \ref{fig0}(b) show the charge radii of Ca isotopes calculated with the meson-exchange effective interaction NL3 and point-coupling effective interaction PC-PK1, respectively. One can find that while smooth curves are obtained with RHB model calculated by both NL3 and PC-PK1 effective interactions, the inverted parabolic-like shapes of Ca isotopes between $N=20$ and $N=28$ are reproduced well after adding the neutron-proton pairing correction terms.
In detail, the quantitative level of charge radii calculated by NL3 is systematically in good agreement with experimental data, but overestimated by PC-PK1 with the RHB* model. 
For Ni isotopes, as shown in Fig.~\ref{fig0}(c) and \ref{fig0}(d), the $N=28$ shell closures are obtained by both the RHB and RHB* models. The RHB* model is better than the RHB model in systematically describing the charge radii of Ni isotopes. For neutron-rich isotopes, there is no kink at $N=32$ or $N=34$ for both isotopes, in consistent with the experimental data~\cite{GarciaRuiz2016_NP12-594,Malbrunot-Ettenauer2022_PRL128-022502}. That is, the magicity of the
neutron numbers $N = 32$ and $N = 34$, which are found from mass measurements~\cite{Wienholtz2013_Nature498-346,Gallant2012_PRL109-032506,Michimasa2018_PRL121-022506}, cannot be observed from the characteristics
of nuclear charge radii along a isotopic chain. }

\begin{figure}[htbp!]
\centering
\includegraphics[width=0.45\textwidth]{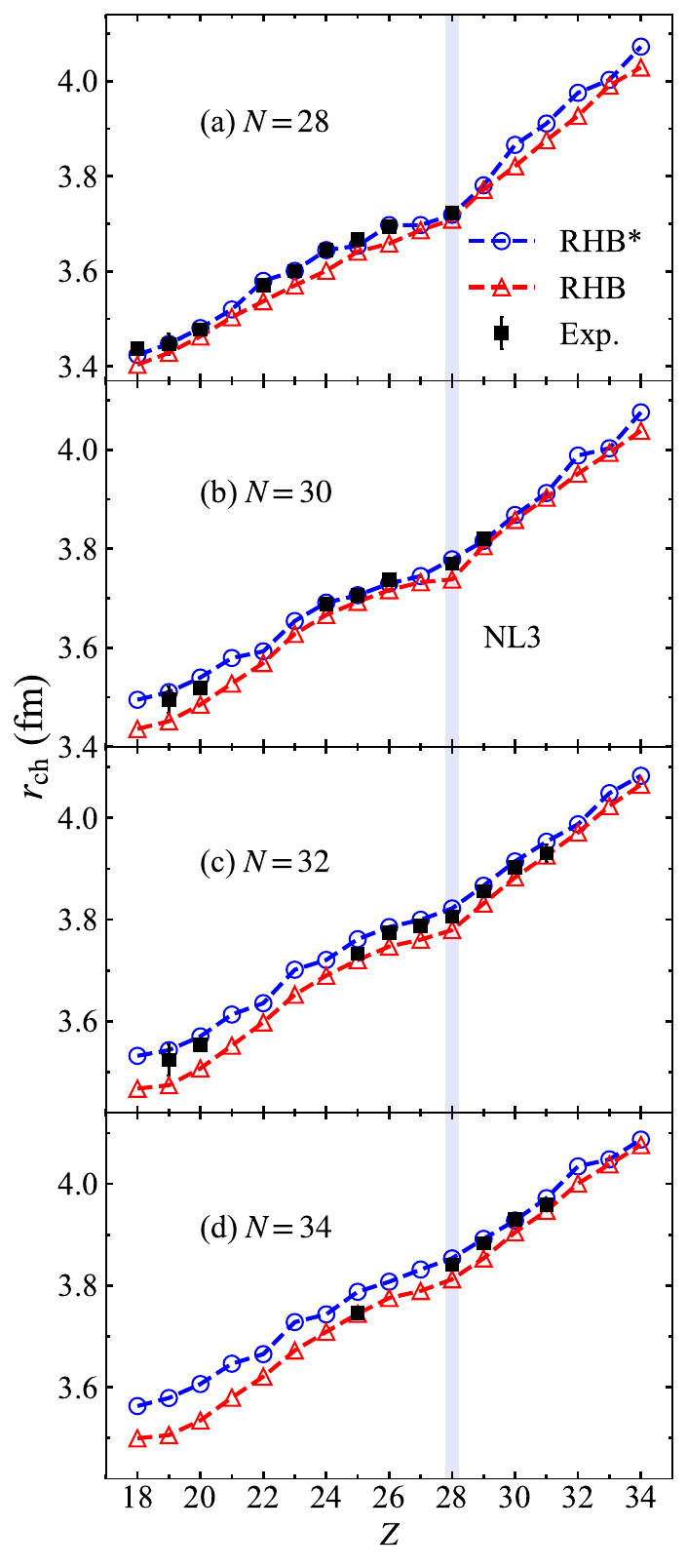}
\caption{(Color online) Charge radii of (a) $N=28$, (b) $30$, (c) $32$, and (d) $34$ isotones calculated by the RHB* (open circle) and RHB (open triangle) models with the NL3 effective interaction. The corresponding experimental data are taken from Refs.~\cite{Angeli2013_ADNDT99-69,Li2021_ADNDT140-101440,Sommer2022_PRL129-132501,Malbrunot-Ettenauer2022_PRL128-022502} (solid square). The gray band represents the proton number $Z=28$.}\label{fig1}
\end{figure}

The shell closure phenomena can be reflected through systematic evolution of charge radii along a long isotonic chain.  
As shown in Ref.~\cite{An2024_PRC109-064302}, the abrupt changes of charge radii across $Z=22$ accur in $N=32$ and $34$ isotonic chains. Meanwhile, the shell quenching effect of $Z=28$ in nuclear charge radii is evidently presented along $N=28$ isotones, but weakened along $N=32$ and $34$ isotonic chains.
To deepen the understanding of these universal and available conclusions, the rms charge radii along $N=28$, $30$, $32$, and $34$ isotones are further investigated based on the MDC-RHB model.
In Fig.~\ref{fig1}, the rms charge radii of $N=28$, $30$, $32$, and $34$ isotones are calculated by the RHB* and RHB models with the meson-exchange effective interaction NL3. As shown in Fig.~\ref{fig1}(a), the RHB* model can better describe the experimental data than its RHB counterpart. The shell quenching effect can be significantly observed across the proton number $Z=28$. 
Meanwhile, the inverted parabolic-like shape of charge radii can be found between the proton numbers $Z=20$ and $Z=28$ along $N=28$ isotones. 
This phenomenon can also be significantly observed between the neutron numbers $N=20$ and $N=28$ in the K and Ca isotopes~\cite{Angeli2013_ADNDT99-69,Li2021_ADNDT140-101440}, but the amplitudes are exaggeratedly enlarged for those nuclei. 
From Figs.~\ref{fig1}(b), \ref{fig1}(c), and~\ref{fig1}(d), the charge radii can also be reproduced well by the RHB* model along $N=30$, $32$, and $34$ isotones, but the value of $^{59}$Mn is visibly overestimated. 
For the RHB model, the calculated results are systematically underestimated against the experimental databases.
Here, it is mentioned that the $Z=28$ shell closure effects are gradually weakened from $N=28$ to $N=34$ isotones in the RHB* model. This is a little different from the results in Ref.~\cite{An2024_PRC109-064302}, in which the zero-range pairing force is used and the $Z=28$ shell closure effect is almost vanished along $N=32$ isotones. 
{Additionally, in Fig.~\ref{fig1}, the abrupt increases in charge radii of $N=30$, $32$, and $34$ isotones are shown across the proton number $Z=22$ after considering the neutron-proton correlations, which is in agreement with results of Ref.~\cite{An2024_PRC109-064302}. We will focus on this phenomenon in detail later.}

\begin{figure}[htbp!]
\centering
\includegraphics[width=0.45\textwidth]{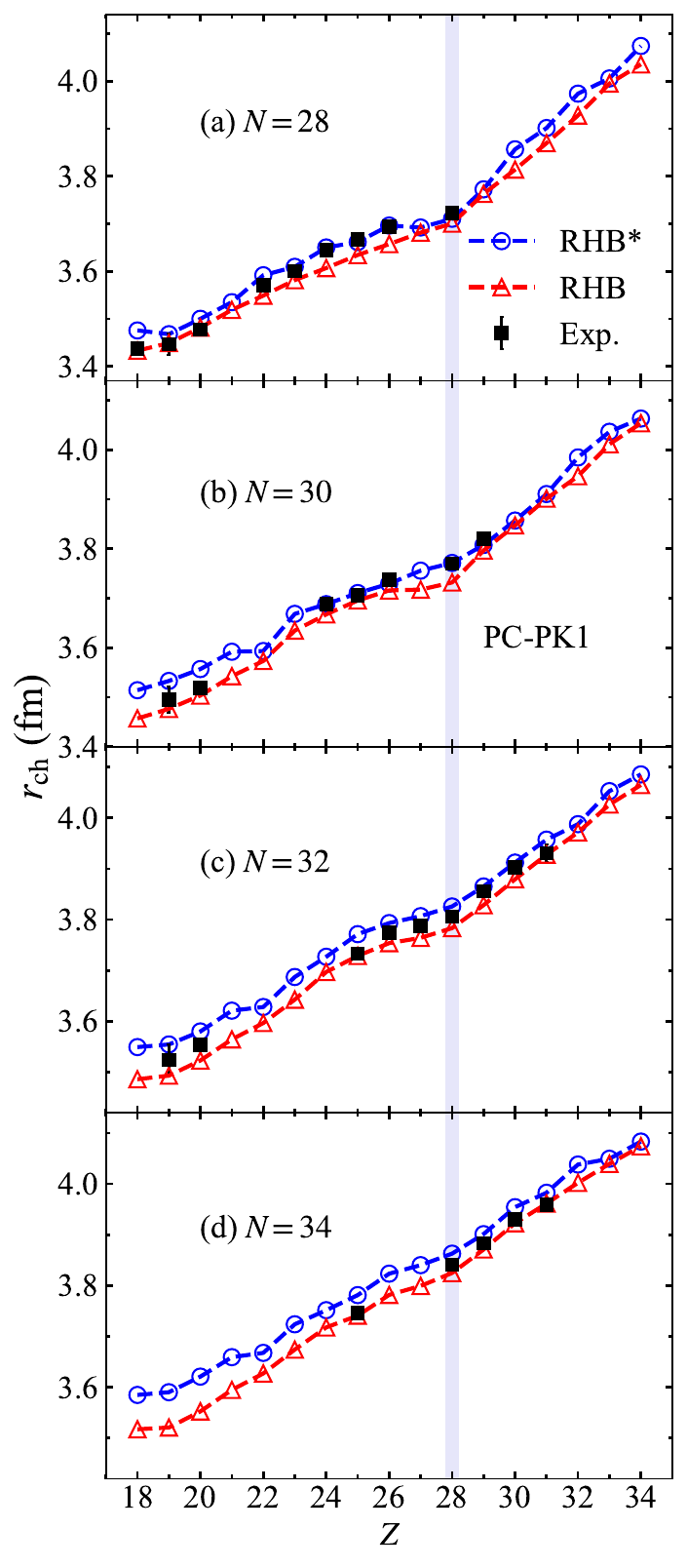}
\caption{(Color online) Same as Fig.~\ref{fig1} but the results obtained by the effective interaction PC-PK1.}\label{fig2}
\end{figure}


In order to clarify the influence of model-dependent on describing the charge radii along $N=28$, $30$, $32$, and $34$ isotones, the point-coupling effective interaction PC-PK1 is also used in this work and the results are shown in Fig.~\ref{fig2}. 
The calculated results show that shell closure effect of $Z=28$ is gradually weakened from $N=28$ to $34$ isotonic chains. This is in accord with those obtained by the NL3 parameter set. Owing to the profound shell effect of $Z=28$ along $N=28$ isotones, the inverted parabolic-like shape of charge radii can also be observed in the RHB* calculations. Toward to proton-rich regions, the results obtained by the RHB* model are slightly overestimated with the PC-PK1 parameter set.
Along $N=30$, $N=32$, and $N=34$ isotonic chains, the increased charge radii data from $Z=22$ to $Z=23$ are about 0.1 fm in the RHB* model. In the RHB calculations, the abruptly increased trend of charge radii is vanished in this region.

\begin{figure}[htbp]
\centering
\includegraphics[width=0.45\textwidth]{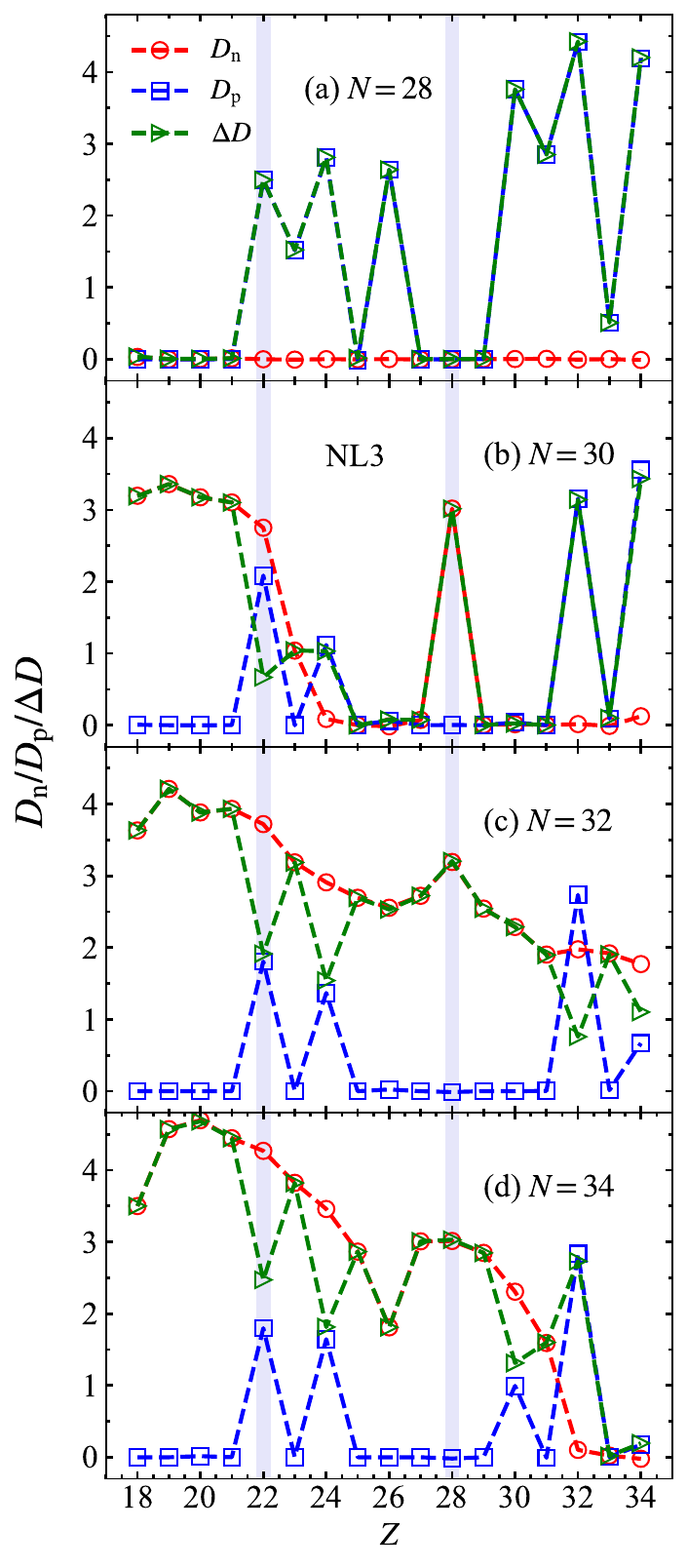}
\caption{(Color online) The $\Delta D$, $D_n$, and $D_p$ values in Eq.~(\ref{eq:r_ch}) for (a) $N=28$, (b) 30, (c) 32, and (d) 34 isotones calculated with the NL3 effective interaction.}\label{fig:DeltaD}
\end{figure}

{Now we discuss the mechanism of the kink appears at $Z=22$ along $N=30$, 32, and 34 isotonic chains. First is the shape evolution. As shown in Refs.~\cite{Lalazissis1996_NPA597-35,Casten1985_PRL54-1991,Togashi2016_PRL117-172502,An2023_CTP75-035301}, the systematic evolution of nuclear charge radii can be influenced by the shape-phase transition. In this work, the deformation parameters for both of $Z=22$ and $Z=23$ isotopes are almost similar along these isotonic chains, namely $\beta_{20}\approx0.20$. This means that the influence of shape deformation is excluded in our discussion. 
Then we check the neutron-proton correction around Fermi surface. For both the charge radii calculated with NL3 and PC-PK1 effective interactions, the kink at $Z=22$ appears after considering the third term in Eq.~(\ref{eq:r_ch}), namely, the neutron-proton correction of the valence particles.
In Fig.~\ref{fig:DeltaD}, we show the values of $\Delta D$, $D_p$, and $D_n$ in Eq.~(\ref{eq:r_ch}) for $N=28$, 30, 32, and 34 isotones. For most of the studied nuclei, the neutron-proton corrections around Fermi surface are nonzero. For $Z=28$, the $\Delta D$ value is zero for the nucleus with $N=28$, but large for those with $N=30$, 32, and 34. This means that $N=28$ and $Z=28$ are magic numbers while $N=32$ and 34 are not. The same picture is found for $Z=20$. From $Z=20$ to $Z=21$, the $\Delta D$ values for all the isotones are similar, indicating the behaviour from $Z=20$ to $Z=21$ for these isotones are almost the same as those in the RHB calculations. At $Z=22$, a sudden increase is obtained for $N=28$. On the contrary, sudden decreases are obtained for $N=30$, 32, and 34 due to the nonzero values of $D_n$ for the corresponding nuclei and the similar $D_p$ values for all the studied isotones. From $Z=22$ to $Z=23$, the $\Delta D$ value increases evidently from $N=28$ to $N=34$ isotonic chains. Therefore, we can conclude that the kink of charge radii at $Z=22$ comes from the sudden decrease of the neuron-proton correction around Fermi surfaces across $Z=22$ for $N=30$, 32, and 34.  
}

\begin{figure}[htbp]
\includegraphics[width=0.45\textwidth]{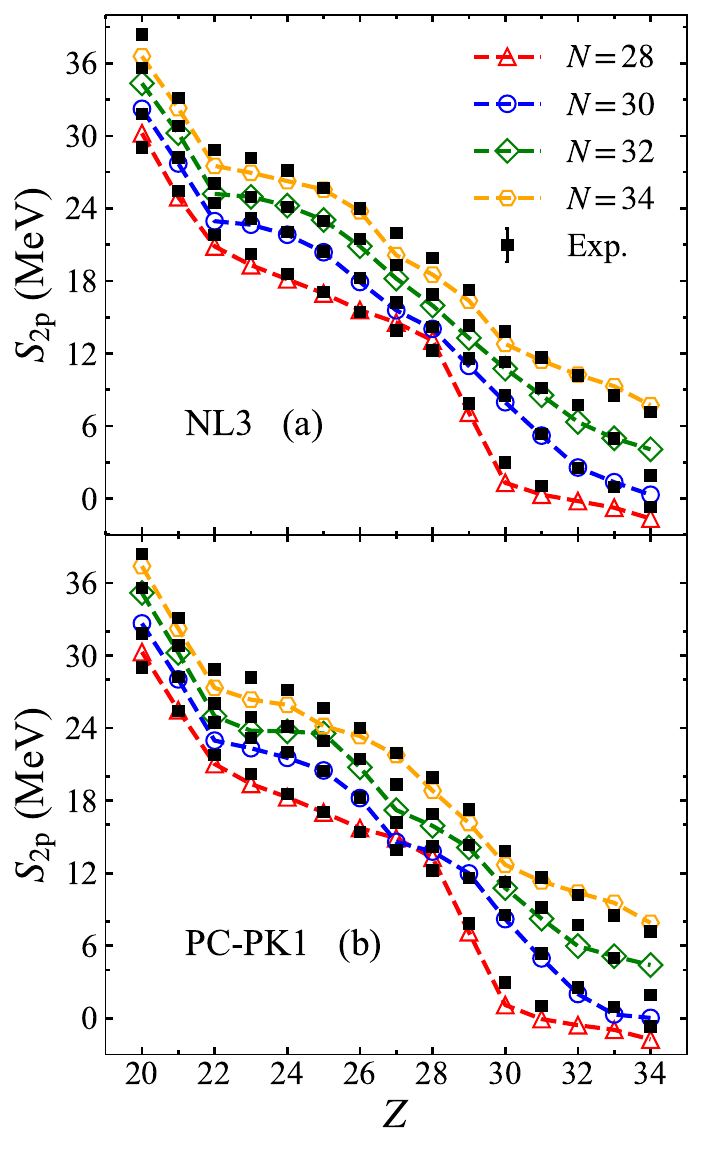}
\caption{(Color online) Two-proton separation energies of $N = 28, 30, 32$, and 34 isotones calculated by the MDC-RHB model with the (a) NL3 and (b) PC-PK1 effective interactions. The corresponding experimental data are taken from Ref.~\cite{Wang2021_ChinPhysC45-030003}.}\label{fig:S2P}
\end{figure}

{The neutron-proton correlations derived from the Casten factor can reproduce the shell closure effect in nuclear charge radii~\cite{Casten1987_PRL58-658,Angeli1991_JPG17-439,Dieperink2009_EPJA42-269,Sheng2015_EPJA51-40}.
For this approach, the isospin interactions deduced from the valence neutrons and protons plays an important role in the validated protocol of matching the experimental charge radii~\cite{Dong2022_PRC105-014308,Dong2023_PLB838-137726}. Moreover, as shown in Refs.~\cite{An2020_PRC102-024307,An2024_PRC109-064302} and this work, the similar approach has been developed through introducing the correlations between the neutrons and protons around Fermi surface.
This neutron-proton correlation can be related to magicity as follow:
When the neutron-proton correlation is decreased significantly with respect to the adjacent counterparts, a kink appears for the charge radii after neutron-proton correlation. So, the nucleus with a kink is more bounded than that without correlation, indicating an enhancement of magicity.
To find the relation between the kink at $Z=22$ and shell quenching, we present the two-proton separation energies $S_{2p}$ of $N=28$, $30$, $32$, and $34$ isotonic chains in Fig.~\ref{fig:S2P}. In this figure, the two-proton separation energies are reproduced well by the MDC-RHB model.
The two-proton separation energies calculated with the NL3 effective interaction are reduce rapidly at $Z=20$ for $N=28$, $30$, $32$, and $34$ isotonic chains and $Z=28$ for $N=28$ isotonic chain, indicating magic numbers $Z=20$ and $Z=28$. However, there is no sudden change at $Z=22$. That is, $Z=22$ is not a magic number in this mass region from the perspective of energy.
For charge radius, the $D_p$ values are similar across $Z=22$ from $N=28$ to $N=34$ (see Fig.~\ref{fig:DeltaD}), meaning the kinks of charge radii at $Z=22$ mainly associate with valence neutrons but not valence protons. Considering mass and decay measurements have manifested that $N=32$ is magic number in neutron-rich K~\cite{Rosenbusch2015_PRL114-202501}, Ca~\cite{Gallant2012_PRL109-032506,Wienholtz2013_Nature498-346}, Sc~\cite{Xu2019_PRC99-064303}, and Ti~\cite{Leistenschneider2018_PRL120-062503} isotopes, and $N=34$ is magic number in Ar~\cite{Liu2019_PRL122-072502}, and Ca~\cite{Michimasa2018_PRL121-022506} isotopes, 
the appearance of the kink may associate with these new neutron magic numbers.

In previous works, the new magic numbers $N=32$ and 34 cannot be found in isotopic chains with $Z\textgreater22$ from precision mass measurements of neutron-rich Ca, Sc, Ti, and V isotopes~\cite{Reiter2018_PRC98-024310,Leistenschneider2021_PRL126-042501,Porter2022_PRC106-024312,Iimura2023_PRL130-012501}, revealing nuclei with $N=32$ and 34 are bounded tightly at the specific isotopic chains until to the $Z=22$ isotopic chain.
Also, in Refs.~\cite{Kortelainen2022_PRC105-L021303,An2024_PRC109-064302}, the charge radii between $N=28$ and $N=40$ exhibit a universal pattern with the increasing mass numbers that is independent of the atomic number along Ca-Zn isotopic chains, indicating that $N=32$ and 34 are not magic numbers in these isotopes from the aspect of charge radii. Unlike those results, our calculations demonstrate the $N=32$ and $N=34$ neutron numbers affect the charge radii after considering neutron-proton correlations in an isotonic chain, which may provide a signature to identify the emergency of magicity.
Since there is no experimental charge radii for $^{52,54,56}$Ti and $^{53,55,57}$V, we suggest to measure the charge radii of them and test the neutron-proton correlations predicted here.
We should emphasize that this is still our guesswork and need further theoretical and experimental verifications.}

Considering the neutron-proton correlations around Fermi surface, the experimental data can be reproduced well by the RHB* model with the effective interactions NL3 and PC-PK1. This provides an available description in predicting the charge radii along $N=28$ isotonic chain. In Table~\ref{tab:1}, charge radii along $N=28$, $30$, $32$, and $34$ isotonic chains obtained by the RHB* model with NL3 and PC-PK1 are listed, respectively. These predicted charge radii should be checked by future experiments.

\begin{table*}[htb!]
\caption{The predicted charge radii data along $N=28,30,32$, and $34$ isotones obtained by the RHB* model with NL3 and PC-PK1 effective interactions are listed, respectively. The experimental data are also shown in the last column and the corresponding systematic error bands are followed in the parentheses (in unit of fm).}\label{tab:1} 
 \doublerulesep 0.1pt \tabcolsep 15pt
	\begin{tabular}{rcccccc}
		\hline
		\hline
		Nucleus~~~ & NL3 &  PC-PK1 & Exp.& NL3 &  PC-PK1 & Exp.\\
		\hline
		{}  & 
\multicolumn{3}{c}{$N=28$}                                                         &\multicolumn{3}{c}{$N=30$}\\
	    Ar$(Z=18)$ &  3.4241 & 3.4757  & 3.4377(44)~\cite{Angeli2013_ADNDT99-69}   &3.4949 & 3.5142 & -\\
        ~K$(Z=19)$ &  3.4473 & 3.4678  & 3.4464(221)~\cite{Li2021_ADNDT140-101440} &3.5103 & 3.5329 & 3.4955(261)~\cite{Li2021_ADNDT140-101440} \\
        Ca$(Z=20)$ &  3.4805 & 3.5001  & 3.4777(25)~\cite{Li2021_ADNDT140-101440}  &3.5398 & 3.5569 & 3.5192(26)~\cite{Li2021_ADNDT140-101440} \\
        Sc$(Z=21)$ &  3.5197 & 3.5351  & -                                         &3.5794 & 3.5920 & -\\
        Ti$(Z=22)$ &  3.5798 & 3.5918  & 3.5704(22)~\cite{Angeli2013_ADNDT99-69}   &3.5927 & 3.5935 & - \\
        ~V$(Z=23)$ &  3.6009 & 3.6095  & 3.6002(22)~\cite{Angeli2013_ADNDT99-69}   &3.6541 & 3.6685 & - \\
        Cr$(Z=24)$ &  3.6445 & 3.6502  & 3.6452(42)~\cite{Angeli2013_ADNDT99-69}   &3.6912 & 3.6882 & 3.6885(74)~\cite{Angeli2013_ADNDT99-69}\\
        Mn$(Z=25)$ &  3.6537 &  3.6616 & 3.6670(37)~\cite{Li2021_ADNDT140-101440}  &3.7062 & 3.7106 & 3.7057(22)~\cite{Li2021_ADNDT140-101440}\\
        Fe$(Z=26)$ &  3.6972 &  3.6963 & 3.6933(19)~\cite{Li2021_ADNDT140-101440}  &3.7299 & 3.7295 & 3.7377(16)~\cite{Angeli2013_ADNDT99-69}  \\
        Co$(Z=27)$ &  3.6973 &  3.6922 & -                                         &3.7450 & 3.7559 &-\\
        Ni$(Z=28)$ &  3.7191 & 3.7105  & {3.7226(03)~\cite{Sommer2022_PRL129-132501}}&3.7788 & 3.7714 & {3.770(2)~\cite{Malbrunot-Ettenauer2022_PRL128-022502}}\\
        Cu$(Z=29)$ &  3.7810 &  3.7725 &-                                          &3.8161 & 3.8078 &3.8200(95)~\cite{Li2021_ADNDT140-101440} \\
        Zn$(Z=30)$ &  3.8660 &  3.8567 &  -                                        &3.8684 & 3.8572 &- \\
        Ga$(Z=31)$ &  3.9109 & 3.9014  & -                                         &3.9128 & 3.9108 &-\\
        Ge$(Z=32)$ &  3.9755 &  3.9738 & -                                         &3.9887 & 3.9848 &- \\
        As$(Z=33)$ &  4.0027 & 4.0058  & -                                         &4.0033 & 4.0367 &-\\
        Se$(Z=34)$ &  4.0727 &  4.0741 & -                                         &4.0760 & 4.0634 &-\\  
       {} &\multicolumn{3}{c}{$N=32$}                                              &\multicolumn{3}{c}{$N=34$}\\
       Ar$(Z=18)$  &  3.5319 & 3.5491  &-                                          &3.5630 & 3.5849 &-\\
       ~K$(Z=19)$  &  3.5431 & 3.5543  &3.5240(313)~\cite{Li2021_ADNDT140-101440}  &3.5789 & 3.5903 &-\\
        Ca$(Z=20)$ &  3.5701 & 3.5799  &3.5531(29)~\cite{Li2021_ADNDT140-101440}   &3.6063 & 3.6209 &-\\
        Sc$(Z=21)$ &  3.6131 & 3.6208  &-                                          &3.6464 & 3.6593 &-\\
        Ti$(Z=22)$ &  3.6355 & 3.6280  &-                                          &3.6652 & 3.6676 &-\\
        ~V$(Z=23)$ &  3.7013 & 3.6870  &-                                          &3.7282 & 3.7238 &-\\
         Cr$(Z=24)$&  3.7206 & 3.7268  &-                                          &3.7434 & 3.7516 &-\\
         Mn$(Z=25)$&  3.7613 & 3.7713  &3.7329(37)~\cite{Li2021_ADNDT140-101440}   &3.7878 & 3.7812 &3.7456(61)~\cite{Li2021_ADNDT140-101440}\\
         Fe$(Z=26)$&  3.7852 & 3.7931  &3.7745(14)~\cite{Angeli2013_ADNDT99-69}    &3.8076 & 3.8235 &-\\
         Co$(Z=27)$&  3.7996 & 3.8067  &3.7875(21)~\cite{Angeli2013_ADNDT99-69}    &3.8316 & 3.8402 &-\\
         Ni$(Z=28)$&  3.8219 & 3.8257  &{3.806(2)~\cite{Malbrunot-Ettenauer2022_PRL128-022502}}   &3.8535 & 3.8629 &{3.835(2)~\cite{Malbrunot-Ettenauer2022_PRL128-022502}}\\
         Cu$(Z=29)$&  3.8665 & 3.8651  &3.8559(55)~\cite{Li2021_ADNDT140-101440}   &3.8919 & 3.9012 &3.8832(26)~\cite{Li2021_ADNDT140-101440}\\
         Zn$(Z=30)$&  3.9144 & 3.9121  &3.9031(69)~\cite{Li2021_ADNDT140-101440}   &3.9280 & 3.9542 &3.9305(47)~\cite{Li2021_ADNDT140-101440}\\
         Ga$(Z=31)$&  3.9531 & 3.9572  &3.9308(174)~\cite{Li2021_ADNDT140-101440}  &3.9721 & 3.9826 &3.9589(26)~\cite{Li2021_ADNDT140-101440}\\
         Ge$(Z=32)$&  3.9874 & 3.9876  & -                                         &4.0345 & 4.0381 &-\\
         As$(Z=33)$&  4.0486 & 4.0526  & -                                         &4.0483 & 4.0495 &-\\
         Se$(Z=34)$&  4.0828 & 4.0855  & -                                         &4.0872 & 4.0833 &-\\       
        
        \hline         
		\hline	
	\end{tabular}
\end{table*}

\begin{figure*}[htbp]
\centering
\includegraphics[width=0.8\textwidth]{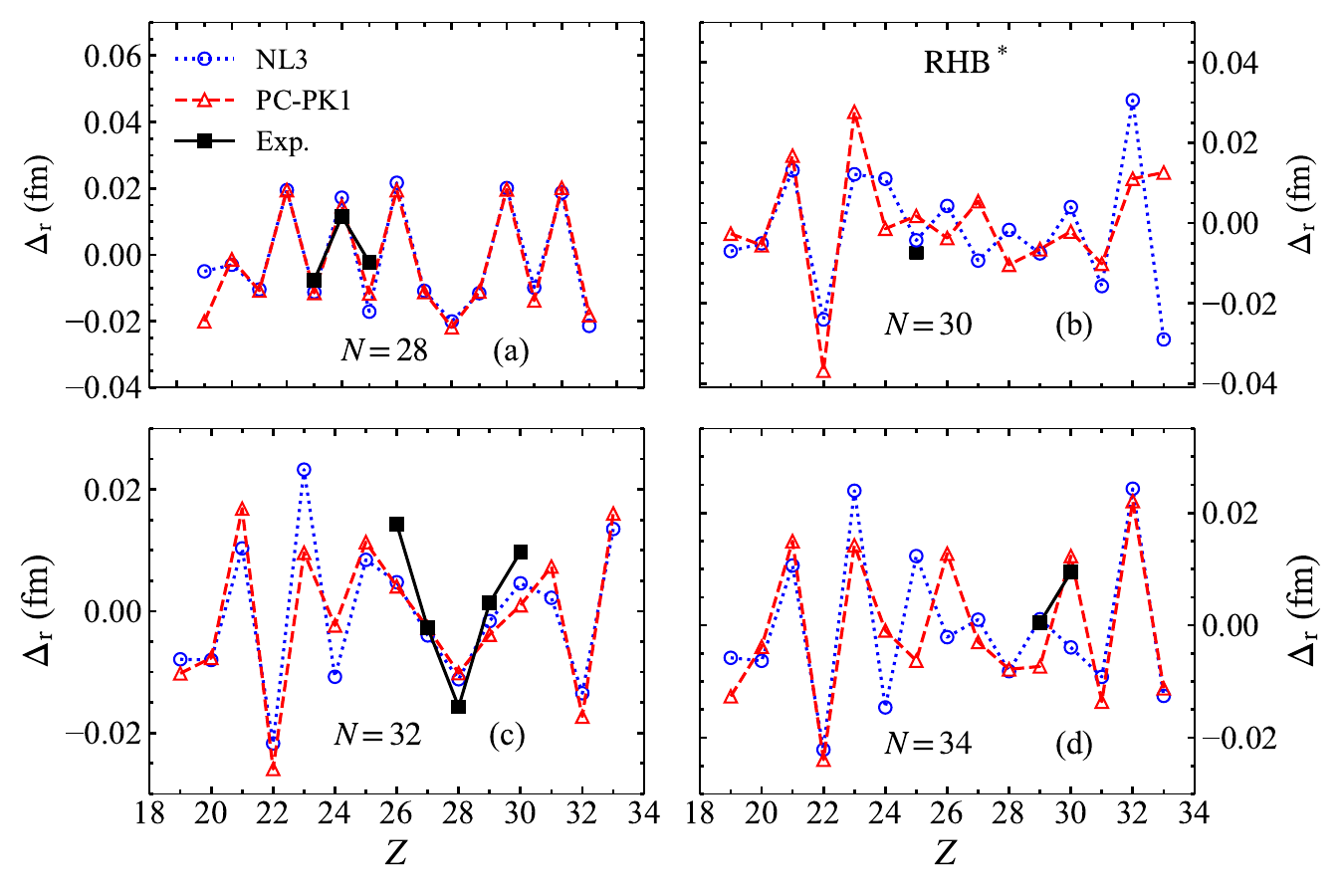}
\caption{(Color online) Odd-even staggerings of charge radii defined by three-point formula as a function of proton numbers along $N=28$, $30$, $32$, and $34$ isotones. The diamond and inverted triangle represent the results obtained by the NL3 and PC-PK1 effective interactions, respectively. The corresponding experimental data are taken from Refs.~\cite{Angeli2013_ADNDT99-69,Li2021_ADNDT140-101440} (solid square).}\label{fig3}
\end{figure*}

To further inspect the local variations of nuclear charge radii along a specific isotonic chain, the three-point OES formula of charge radii is recalled as follows~\cite{Reinhard2017_PRC95-064328}:
\begin{equation}
 \Delta_r(N,Z)=\dfrac{1}{2}[2r_{\rm ch}(N,Z)-r_{\rm ch}(N,Z-1)-r_{\rm ch}(N,Z+1)],
\end{equation}
where $r_{\rm ch}(N,Z)$ is the rms charge radii of a nucleus with neutron number $N$ and proton number $Z$.
In Fig.~\ref{fig3}, the OES of charge radii along $N=28$, $30$, $32$, and $34$ isotones as a function of proton numbers are plotted. Along $N=28$ isotones, the results obtained by both NL3 and PC-PK1 effective interactions reproduce the experimental data reasonablely. At the proton number $Z=19$, a deviation can be observed between these two effective interactions. For $N=30$ isotones, the opposite trend can be distinguished from $Z=25$ to $Z=29$ for these two interactions. The same scenario can also be found at the proton number $Z=33$.

Along $N=32$ isotones, results derived from both NL3 and PC-PK1 interactions reproduce the experimental data,
but a slightly deviation occurs between the proton numbers $Z=26$ and $Z=30$. 
For $N=34$ isotones, the sign of the OES in charge radii calculated by NL3 is opposite from those calculated by PC-PK1 from $Z=25$ to $Z=29$ except for the case of $Z=28$. The experimental data are just reproduced by the NL3 interaction, while the $\Delta_r$ at $Z=30$ is poorly described by the PC-PK1 effective interaction. 
Besides, as shown in Figs.~\ref{fig3}(a), ~\ref{fig3}(c), and~\ref{fig3}(d), the odd-even staggering effect is weakened at the proton number $Z=28$ for both of NL3 and PC-PK1 effective interactions. 
This phenomenon, namely the weakened OES in charge radii, can be generally observed at the neutron numbers $N=28$, $50$, $82$, and $126$~\cite{An2020_PRC102-024307,An2022_PRC105-014325}. 
While in Fig.~\ref{fig3}(b), the local variations at the proton number $Z=28$ are ambiguity for PC-PK1  and NL3 effective interactions.
Furthermore, the amplitudes of OES in charge radii are enlarged at the proton number $Z=22$ in $N=30$, $32$, and $34$ isotones from the RHB* model,  manifesting the abruptly increased trend in this mass region.


\section{SUMMARY}\label{sec4}

{Charge radii of Ca and Ni isotopes, and $N=28$, $30$, $32$, and $34$ isotones are calculated based on the MDC-RHB model.} The separable pairing force of finite range is used to solve the RHB equation. The neutron-proton correlations around Fermi surface are taken into account properly in describing the systematic trend of changes of nuclear charge radii, namely the extended RHB* model. 
To further inspect the universal aspects of nuclear charge radii, the generally used meson-exchange and point-coupling effective interactions are recalled in our calculations, respectively. 

{The results obtained by the RHB* model reproduce the experimental charge radii of the studied nuclei well, pointing out that the neutron-proton correlations around Fermi surface seem to play an indispensable role in describing the systematic evolution of nuclear charge radii across this mass region. The charge radii of Ca and Ni
isotopes are improved in comparison with experimental data after neutron-proton pairing corrections. The magicity of the neutron numbers $N=32$ and $N=34$ cannot be observed from the characteristics of nuclear charge radii along a isotopic chain, in consistent with the experimental data.}
For isotonic chain calculations, the shell closure effect of $Z=28$ is sequentially weakened from the $N=28$ to $N=34$ isotones. Furthermore, the inverted parabolic-like shape of charge radii in $N=28$ isotones can be clearly observed between the proton numbers $Z=20$ and $Z=28$. But for $N=30$, $32$, and $34$ isotones, this trend is still sequentially reduced. The remarkably raise of charge radii along $N=30$, $32$, and $34$ isotonic chains can be found from the proton number $Z=22$ to $Z=23$ after considering the neutron-proton corrections. {By analysing the neutron-proton correlation term in our framework, we find the kink at $Z=22$ comes from the sudden decrease of the neuron-proton correlation around Fermi surfaces, in which the valence neutrons play an important role.} Combining the existing literature, we can further identify that the emergence of new magicity for $N=32$ and $34$ may be characterized through the universal trend of changes of nuclear charge radii. In addition, the available values of charge radii along $N=28$, $30$, $32$, and $34$ isotonic chains are predicted by the RHB* model. This may be a useful guideline in experiments. Therefore, more charge radii data are urgently required in this mass region.
	
\begin{acknowledgments}
This work has been supported by the National Natural Science Foundation of China (Grant No. 12205057), the Central Government Guides Local Scientific and Technological Development Fund Projects (Grant No. Guike ZY22096024), the Science and Technology Plan Project of Guangxi (Grant No. Guike AD23026250), and the Natural Science Foundation of Guangxi (Grant No. 2023GXNSFAA026016). R. A. is grateful for the support of the Open Project of Guangxi Key Laboratory of Nuclear Physics and Nuclear Technology (Grant No. NLK2023-05).
\end{acknowledgments}
	
\bibliographystyle{apsrev4-2}
\bibliography{refsanw.bib}
	
\end{document}